\documentclass[iop]{emulateapj}

\makeatletter
\newcommand\tsb[1]{\@textsubscript{\selectfont#1}}
\def\@textsubscript#1{{\m@th\ensuremath{_{\mbox{\fontsize\sf@size\z@#1}}}}}
\newcommand\tsp[1]{\@textsuperscript{\selectfont#1}}
\def\@textsuperscript#1{{\m@th\ensuremath{^{\mbox{\fontsize\sf@size\z@#1}}}}}

\providecommand{\e}[1]{\ensuremath{\times 10^{#1}}}
\providecommand{\lsr}[1]{\textit{V}\tsb{LSR}}

\providecommand{\rsolar}[1]{#1 R\tsb{$\odot$}}
\providecommand{\kms}{km s\tsp{-1}}

\providecommand{\cm}{cm\tsp{-3}}
\providecommand{\ndot}[1]{$\mathsf{\dot{N}_{#1}}$}

\usepackage{graphicx,color}



\begin{document}

\title{EUV and X-Ray Observations of Comet Lovejoy (C/2011 W3) in the Lower Corona}
    
\author{Patrick I. McCauley\tsp{1}, Steven H. Saar\tsp{1}, John C. Raymond\tsp{1}, Yuan-Kuen Ko\tsp{2}, and Pascal Saint-Hilaire\tsp{3}}
\affil{\tsp{1}Harvard-Smithsonian Center for Astrophysics, 60 Garden St, Cambridge, MA 02138 \\ 
	\tsp{2}Space Science Division, Naval Research Laboratory, 4555 Overlook Ave., SW, Washington, DC 20375 \\
	\tsp{3}Space Sciences Laboratory, University of California, Berkeley, CA 94720}
\email{pmccauley@cfa.harvard.edu}
	
\shorttitle{EUV \& X-Ray Observations of C/2011 W3}
\shortauthors{McCauley et al.}

    
\begin{abstract}

We present an analysis of EUV and soft X-ray emission detected toward Comet Lovejoy 
(C/2011 W3) during its post-perihelion traverse of the solar corona on December 16, 
2011. Observations were recorded by the Atmospheric Imaging Assembly (AIA)
aboard the Solar Dynamics Observatory and the X-Ray Telescope (XRT) aboard 
Hinode. A single set of contemporaneous images is explored in detail, along with prefatory 
consideration for time evolution using only the 171 \AA{} data. For each of the 
eight passbands, we characterize the emission and derive outgassing rates where applicable. 
As material sublimates from the nucleus and is immersed 
in coronal plasma, it rapidly ionizes through charge states seldom seen in this environment. The AIA 
data show four stages of oxygen ionization (O III - O VI) along with 
C IV, while XRT likely captured emission from O VII, a line typical of the corona. With a nucleus of at least 
several hundred meters upon approach to a perihelion that brought the comet to within 0.2 \rsolar{} of 
the photosphere, Lovejoy was the most significant sungrazer in recent history. Correspondingly high
outgassing rates on the order of 10\tsp{32.5} oxygen atoms per second are estimated. Assuming that 
the neutral oxygen comes from water, this translates to a mass-loss rate of $\sim$9.5\e{9} g s\tsp{-1}, and 
based only on the 171 \AA{} observations, we find a total mass loss of $\sim$10\tsp{13} g over the AIA egress.
Additional and supporting analyses 
include a differential emission measure to characterize the coronal environment, consideration for the 
opening angle, and a comparison of the emission's leading edge with the expected position of the nucleus.

\end{abstract}



\section{Introduction}
\label{intro}


  \begin{figure*}[t]
 \epsscale{1}
 \plotone{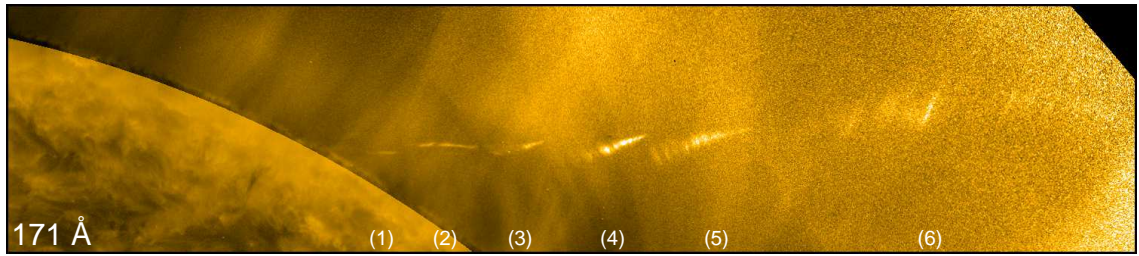}
 \caption{171 \AA{} AIA images at 6 times: (1) 00:41:36 UT (2) 00:43:00 (3) 00:44:12 (4) 00:46:12 
(5) 00:48:12 (6) 00:56:00. We focus this work primarily on multi-wavelength observations at position (4), with some consideration for time 
evolution in \S\ref{time}. These images were produced using 
a radial filter; see our \href{http://aia.cfa.harvard.edu/movies/comet_lovejoy_egress_171.mp4}{online material for the corresponding movie}.}
 \label{fig:rfilter}
 \end{figure*}

Comet Lovejoy (C/2011 W3) is the most significant sungrazer since the launch of the Solar Dynamics Observatory 
(SDO; \citealt{Pe12}) and the first to be detected by the X-Ray Telescope (XRT; \citealt{Go07}) 
aboard Hinode \citep{Ko07}. Its perihelion passage brought the comet to within just \rsolar{0.2} of the 
photosphere, providing an unprecedented 
glimpse of a large comet immersed in the lower corona. Lovejoy is a member of the Kreutz family of 
sungrazing comets, which account for a majority of all sungrazers and are observed in large numbers 
each year by the Solar and Heliospheric Observatory (SOHO; \citealt{Do95}). SOHO's Large Angle and Spectrometric 
Coronagraph (LASCO; \citealt{Br95}) has proven a superb discoverer of these objects \citep{Bi02}, having 
imaged over 1800, and the UltraViolet Coronagraph Spectrometer (UVCS; \citealt{Ko95}) has returned 
spectra from several Kreutz members \citep{Ra98, Be07}.

The group is thought 
to have been formed from the successive fragmentation of a single progenitor as little as 2500 years ago 
\citep{SC02, SC04, SC07} and is principally comprised of 
meter-sized objects that are most often destroyed well before perihelion \citep{Kn10}. Until Lovejoy, none 
of the  Kreutz members observed by SOHO have survived to emerge from behind the occulting disk of the 
coronagraph \citep{Is02}. A few have made it into the lower corona, but it was not until the launch of SDO in 
2010 that they could be followed using the Atmospheric Imaging Assembly (AIA; \citealt{Le12}). \citet{Sc12} 
reported the first such observation, which witnessed the destruction of Kreutz fragment C/2011 N3 in July of 
2011 about midway through its transit of the solar disk. When Lovejoy followed five months later, it seemed 
that it would likely meet the same fate, this time behind the Sun from AIA's perspective. 
Instead, the comet emerged from behind the limb just moments after SDO 
had finished slewing to its usual disk-center pointing after being off-pointed to observe the ingress.


Exactly how a comet like C/2011 W3 is able to survive its closest approach is still not entirely clear, largely
 because there is uncertainty regarding its size and composition. Based on the visible magnitude well before perihelion, 
 the nucleus is expected to have been less than $\sim$1 km in diameter \citep{Gu12}. Preliminary 
 estimates from SOHO/UVCS suggest a diameter of 400 m upon reaching \rsolar{6.8} during the ingress \citep{Ra13}. 
 Finally, after having seemingly escaped, the nucleus was destroyed at $\sim$\rsolar{31}, 1.6 days past perihelion 
 \citep{Se12}. From mass lost arguments using the visible brightness of the dust tail, \citet{SC12} estimate that the 
 diameter of the nucleus was on order 200 m at the time of its destruction. If the comet's makeup was consistent with 
 the canonical ``rubble pile" model, then the tidal forces within the Sun's Roche lobe should have torn it apart 
 opposed only to self gravity and a meager tensile strength. \citet{Gu12} explore this possibility and suggest that 
 the cohesive pressure exerted by near-isotropic outgassing could be a sufficient counter to the tidal forces. 

However, \citet{SC12} argue that the delayed destruction of the nucleus is evidence that the comet was no ``rubble pile" 
and possessed tensile strength appreciable enough to remain structurally unperturbed by tidal forces. They suggest that 
the thermal stresses experienced as the comet swept into the lower corona would have taken time to propagate into 
the interior, after which pockets of water ice would have exploded upon reaching $\sim$130 K due to an exothermic 
reaction related to the crystallization of amorphous ice \citep{Sc89}. In this scenario, the 
comet is destroyed by successive, large fragmentation events rather than steady outgassing. \citet{SC12} include a 
discussion of preliminary results from the work presented here, which they explain in the context of their fragmentation 
model. We will revisit this in \S\ref{outbursts}.


C/2011 W3 is the first comet to have been detected by a solar X-ray imager, but it is not the first comet to exhibit X-rays 
at all. Sixteen years ago, \citet{Li96} reported the puzzling detection of extreme ultraviolet (EUV) and X-ray emission from comet C/1996 
B2 (Hyakutake) using the X-Ray Telescope aboard ROSAT. Shortly thereafter, such emission was found to emanate 
from any comet within 3 AU \citep{De97}. \citet{Cr97} and \citet{Kr97} explained that this high-energy emission 
arose from charge exchange between solar wind ions and the neutral gas of cometary atmospheres, which are much 
too cold to produce such energetic photons on their own. But any neutral species are quickly ionized within the AIA 
and XRT fields of view, so a different mechanism is needed to explain the emission described here.

In their work on the aforementioned C/2011 N3, \citet{Sc12} noted that the EUV emission could be explained by 
ionization and excitation states experienced as shed cometary material equilibrated with coronal plasma. \citet{BP12} 
expanded upon this by providing a detailed summary of the means by which sublimated cometary molecules are 
destroyed and ionized. They constructed a time-dependent model of the cometary emission using a simple geometry and
determined which ionization states of which elements would contribute to AIA's EUV channels, noting that these 
differ significantly from the lines typically important to AIA observations.

We adopt the scenario proposed by \citet{Sc12} and \citet{BP12} and assume that the EUV and X-ray emission arises from 
the ionization of cometary material after it is immersed in the corona. Sublimated molecules flow into the coma 
at velocities of at least a few \kms{} \citep{Co00} and are swiftly photodissociated into their atomic constituents. 
We make the simplifying assumption that 
water dominates the composition of this material, and since hydrogen does not contribute significantly to any of our 
passbands, our observations can be largely characterized by which 
charge states of oxygen contribute most to the various channels. 
We identify the dominant lines in each of our passbands in \S\ref{observations}, which also includes a more thorough 
discussion of the \citet{BP12} model and how our results compare to theirs.


 \begin{figure*}[t]
 \epsscale{1}
 \plotone{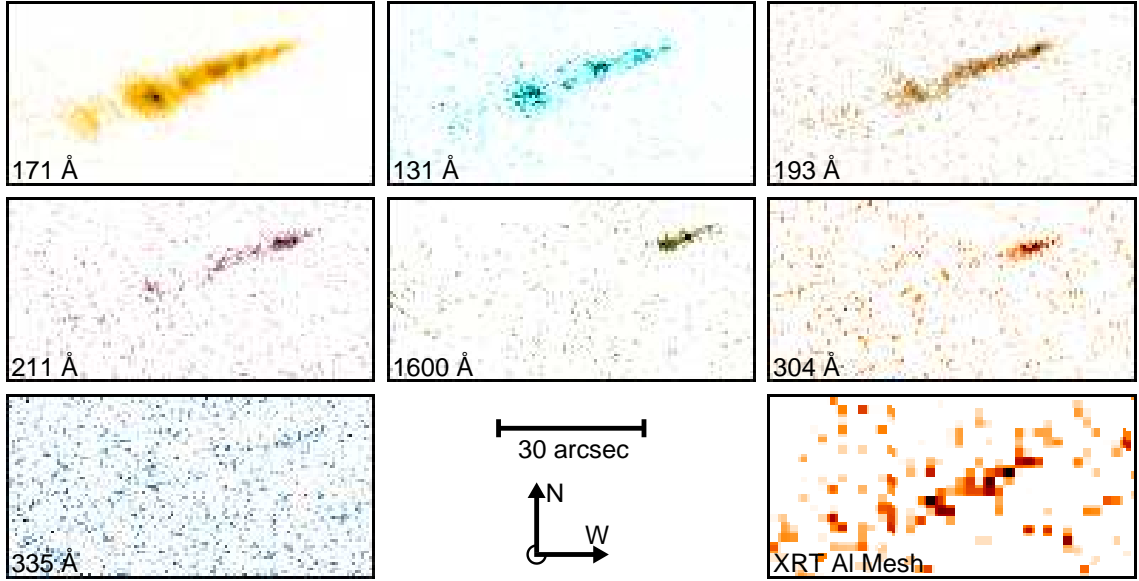}
 \caption{Background subtracted images of the cometary emission at 00:46:12 UT on 2012/12/16.  This 
 was $\sim$30 min post-perihelion, corresponding to a height of about \rsolar{0.56} ($\sim$320,000 km) above the photosphere.
 Note that the color tables are inverted, so the darkest regions are those of highest intensity.}
 \label{fig:images}
 \end{figure*}


This paper is structured as follows: \S\ref{observations} describes our observations, the means by which cometary 
flux was extracted from the coronal background, and the source of EUV and X-ray photons. In \S\ref{dem}, we use a 
differential emission measure to probe the coronal environment, and \S\ref{outgassing} details our calculations of 
outgassing rates from the observed fluxes. The foremost extent of the emission in each channel and the expected 
position of the nucleus are given in \S\ref{positions}. \S\ref{angle} includes consideration for the 171 \AA{} opening angle, 
and \S\ref{time} examines time evolution over the egress again using the 171 \AA{} observations. We discuss our work 
in the context of others in \S\ref{discussion}. This includes commentary on the \citet{SC12} outburst model,  
the source of neutral oxygen (water ice vs. dust grains), how our results compare with those of \citet{Sc12} for comet 
C/2011 N3, and what our results imply about the size of the nucleus. 
Finally, \S\ref{conclusion} summarizes our findings.


\section{Observations}
\label{observations}
	
The observations reported here were taken on 
December 16, 2011 during the egress of Comet Lovejoy, after it had emerged from behind the solar disk having survived 
perihelion. Figure~\ref{fig:rfilter} provides an overview of our dataset by overlaying six 171 \AA{} AIA images at different times.  
These images were produced using a radial filter\footnote{The radial filter divides the off-limb component of an AIA image 
into concentric rings and scales each ring as a function its radius 
and brightness relative to neighboring rings. As such, flux is not conserved. The brightness of each pixel corresponds 
only to its intensity relative to pixels of the same radius. Source code is available in SolarSoft under $<$aia\_rfilter$>$.}, which 
allows us to see the development of both cometary and coronal emission in one 
coherent set of images. See our online material for the corresponding movie.

SDO/AIA records images with a 12 s cadence in seven narrowband EUV channels (94, 131, 171, 193, 211, 304, \& 
335 \AA{}) and with a lower cadence in 3 UV bands (1600, 1700, \& 4500 \AA{}). The resolution is 0.6\arcsec per pixel 
over a 4096$\times$4096 px CCD. 
Detections were made in each channel with the exceptions of the 94, 1700, and 4500 \AA{} bands. Hinode/XRT made 
successful observations using its thinnest soft X-ray filter, Al-mesh, and recorded non-detections using two thicker filters 
(Be-med \& Al-med). These observations had a cadence of 7 s and a resolution of 2.06\arcsec per pixel. The dominant 
ions contributing to each channel are listed in Table~\ref{tab:ions}, and discussion of how we arrived at these is 
given in the next two sections. 

Our analysis focuses primarily on a single set of images centered around 
00:46:12 UT (position 4 of Fig~\ref{fig:rfilter}), which is about 30 minutes after perihelion and corresponds to 
a height above the photosphere of about \rsolar{0.56} ($\sim$320,000 km). This height and those 
quoted later are determined 
from the most recent ephemeris solution; see \S\ref{positions} for additional discussion of the 
expected trajectory. We chose this time for a strong XRT signal and maximum 
temporal alignment between XRT and the AIA/171 channel, which was important 
for characterizing the X-ray production (see \S\ref{xrt}). The thin XRT filters are 
also affected by CCD contamination spots that attenuate longer X-ray wavelengths, and so 
this time was also chosen because the location of the comet is free of contamination. 

There are a number of interesting features and morphological developments to note as the comet 
progresses through the AIA field-of-view (FOV) during the egress. These include apparent interactions 
with the coronal magnetic field (\textbf{B}) and spikes in the measured intensity. We will revisit these topics in \S\ref{time}, 
where we consider time evolution using only the 171 \AA{} observations.


\subsection{Background Subtraction}
\label{background}

Background subtraction was necessary to separate cometary and coronal emission. 
The backgrounds were constructed from averages of 10 frames, 5 before and 5 after 
the comet appeared within a $\sim$100\arcsec{} region surrounding the emission. 
Before subtraction, the backgrounds were smoothed using a 5x5 pixel boxcar average to 
limit the introduction of additional noise. Minor over- and under-subtractions 
related to intensity gradients across the background frames were accounted 
for by removing a subsequent linear fit to the column and row averages outside the comet region. 
Background subtracted images for each channel are displayed in  Figure~\ref{fig:images}.

To extract counts from the comet without including unnecessary background pixels, the flux
was taken from a region contoured to the shape of the emission. This was done by 
thresholding the subtracted images, selecting contiguous pixels, and dilating the resulting 
mask until it comfortably included all observed emission. Columns 4 and 5 of Table~\ref{tab:flux} 
list the results of this. Column 4 includes fluxes integrated over a single, all-encompassing (global) mask,
while the values in column 5 were derived from masks individually contoured to the emission 
in each channel. Note that if the background subtraction were perfect, columns 4 and 5 would be 
equivalent. 

The quoted errors reflect statistical uncertainties based on the total counts recorded for 
comet and background, along with an additional term 
that attempts to quantify error from the background removal using the RMS noise outside of the 
comet region after background subtraction. Columns 6 and 7 list the percentage of the total flux that is attributed 
to the comet using the global and individual masks, respectively. 
Note that the difference between these two values is reflective of both the extent of the cometary emission 
in a given channel and the intensity of the background. 
Values for the cometary portion of the total flux range between just 3\% for the XRT observation and 43\% for AIA/131 \AA{}.


\begin{deluxetable}{l c c c c}
\tabletypesize{\scriptsize}
\tablecolumns{5} 
\tablewidth{0pc} 
\tablecaption{Dominant Emission Contributors} 
\tablehead{ 
\colhead{Ion} & 
\colhead{Peak\tablenotemark{a}} & 
\colhead{FWHM\tablenotemark{a}} & 
\colhead{Ionization\tablenotemark{b}} &
\colhead{Channel\tablenotemark{c}}
\\ 
\colhead{} & 
\colhead{log(T)} & 
\colhead{log(T)} & 
\colhead{Rate q\tsb{i}}&
\colhead{}
\\ 
\colhead{} & 
\colhead{(K)} & 
\colhead{(K)} & 
\colhead{(cm\tsp{-3} s\tsp{-1})}&
\colhead{(\AA{})}
}
\startdata
C IV & 5.05 & 0.35 & 2.33\e{-9} & 1600 \\
O III & 5.1 & 0.47 & 1.32\e{-8} &  \textbf{304}, 335 \\
O IV & 5.3 & 0.35 & 5.38\e{-9} & 211, 304 \\
O V & 5.4 & 0.35 & 1.83\e{-9} & \textbf{171}, 193  \\
O VI & 5.55 & 0.35 & 6.57\e{-10} & 131, 171, \textbf{335} \\
O VII & 6.3 & 0.47 & \nodata & Al-Mesh\tablenotemark{d}
\enddata
\tablenotetext{a}{Temperatures (peak and FWHM range) are based on the G(T) distribution 
functions from CHIANTI, which represent the temperatures these ions would be associated with under equilibrium conditions.
These have been provided for context but are not to be directly associated with the comet observations 
because the cometary plasma is not in ionization equilibrium.}
\tablenotetext{b}{q\tsb{i} calculated for log(T) = 6.15 K. Rates for O I and O II are 
8.04\e{-8} and 2.88\e{-8} cm\tsp{-3} s\tsp{-1}, respectively. See \citet{BP12} for a detailed discussion on ionization rates.}
\tablenotetext{c}{Boldfaced font indicates the more dominant ion for those channels listed twice.}
\tablenotetext{d}{O VII emits near the peak sensitivity of XRT's Al-mesh filter at 22.1 \& 21.6 \AA. 
See Fig~\ref{fig:xrtres} for the Al-mesh wavelength response.}
\label{tab:ions}
\end{deluxetable}


\subsection{EUV Emission}
\label{euv}

As described in \S\ref{intro}, the EUV emission observed by AIA likely arises from the ionization of shed 
cometary material as it equilibrates with coronal plasma.
This material originates from the neutral 
environment of the nucleus, and thus low ionization states not typically observed by AIA must be considered. 
We have assumed that the bulk of the sublimated material is water, and since the soon-ionized hydrogen 
does not contribute significantly to any of the AIA bands, oxygen is the most important atom to consider 
(see \S\ref{oxygen} for additional discussion on the source of neutral oxygen). 

To this end, we have 
used version 6 of the CHIANTI\footnote{CHIANTI is a collaborative project involving George Mason University, 
the University of Michigan (USA) and the University of Cambridge (UK).}
spectroscopic database \citep{chianti,chianti6}, coupled with the effective areas of the AIA filters, to determine 
which O lines dominate each channel. These results are listed in Table~\ref{tab:ions}, and some additional 
discussion of the emission process is given in \S\ref{outgassing}. Note that there are no significant O lines 
covered by the 1600 \AA{} UV filter, and for this we expect that the emission arises principally from C IV. 

Also note that morphological differences between the bands, specifically the length of the tail, result mainly from differences 
in the ionization times. 
As we will describe in \S\ref{outgassing}, a neutral O atom is stripped of its first electron in about one 
tenth of a second after being dropped into the coronal plasma. The ionization rate of 
each subsequent state decreases by a factor of 3--5, and so 
channels dominated by low charge states (1600 \& 304 \AA{}) exhibit shorter 
tails because these states do not persist for long before ionizing into the next stage. 
Similarly, the intermediate, O IV image (211 \AA{}) shows an intermediate extent, 
and likewise for the higher ionization states.  

The O VI emission exhibited by the 
171 and 131 \AA{} bands will persist until the atoms progress to O VII or the comet-related densities have 
sufficiently diffused into the coronal background. The next section will show that XRT likely sees material that has indeed 
made it to O VII, but as will be described in \S\ref{131335}, we suspect that this occurs for only about 1/3 of the O VI atoms. 
Examination of the O VI images suggests that the cometary material becomes indistinguishable from the background 
after about two minutes near 00:46:12 UT. This time is greatly extended further along in the dataset, likely due to 
a lower coronal density; see \S\ref{morph} for additional details. 


A more complete assessment of the cometary emission observed by AIA has been carried out by \citet{BP12}, who
conducted a full, time-dependent analysis with a simple geometry. The coma was modeled as a cylinder 
populated by concentric shells of successive ionic species, which 
form as material from the nucleus is sublimated and ionized. Once immersed in the corona, cometary atoms 
may be ionized in several ways; \citeauthor{BP12} include considerations for charge exchange, electron- 
and proton-impact, and photoionization. Each of these effects is examined for the most abundant  
elements determined for Comet 1P/Halley by \citet{De88}, which include H, O, C, N, Si, Fe, Mg, and S (listed 
here in order of descending abundance). The lines relevant to AIA are identified and displayed in a series 
of figures showing their emission versus typical quiet-Sun conditions. 

\citet{BP12} find that the emission in most channels is dominated by 
oxygen, and the O ions they identify are consistent with what we present in Table~\ref{tab:ions} based 
on our more simplistic approach. The only exception to this is the 335 \AA{} band, for which we 
find the O VI lines near 131 \AA{} to dominate over the O III lines near 335 \AA{}. We will discuss 
this further in \S\ref{131335}. \citeauthor{BP12} find no significant emission from H, C, N, Si, Mg, or S. 
They do, however, find that iron contributes significantly to the 171 \AA{} (by $\sim$40\%) and 131 \AA{} ($\sim$25\%) channels. 
(The 94 \AA{} channel is also found to be dominated by Fe, but a reliable signal was not detected at this wavelength.) The 
relative contributions of O and Fe in their model depend somewhat on
outflow velocity, for which 17 \kms{} is used based on observations of C/2011 N3. 
This dependence is characterized in Figure 10 of their paper and may also be important to our discussion 
on the source of neutral oxygen in \S\ref{oxygen}.


 \subsection{Soft X-Ray Emission}
\label{xrt}

\begin{figure}[t]
 \epsscale{1}
 \plotone{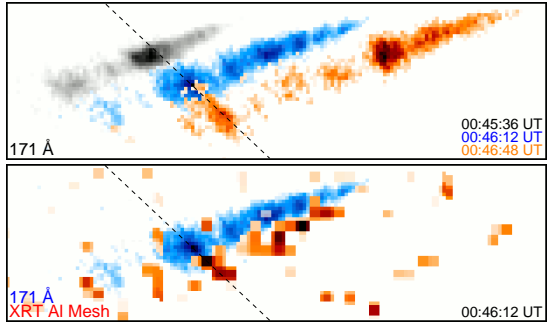}
 \caption{\textit{Top}: 3 successive AIA/171 images, spaced 36 s apart, that demonstrate the direction 
 ionized cometary material travels along the local, coronal \textbf{B} field. \textit{Bottom}: Overlay of  
 simultaneous XRT and AIA observations. The XRT emission is offset along the direction established above,  
 suggesting that XRT samples more highly ionized material that has moved farther down the field lines.}
 \label{fig:overlay}
 \end{figure}

Careful alignment reveals the XRT emission to be offset from AIA. As outgassed material ionizes, 
its motion soon becomes dominated by the local magnetic 
field. Newly formed plasma from the comet can be observed drifting along field lines in various 
directions throughout the AIA dataset (see \S\ref{morph} for additional details). At 00:46:12 UT, this motion 
is in a southwest direction that matches the observed offset between nearly simultaneous AIA and 
XRT images. This is illustrated by Figure~\ref{fig:overlay}. The upper panel shows three successive 
AIA/171 images to establish the direction of motion along the \textbf{B} field, while the lower panel shows the 
XRT offset. The AIA and XRT images in the lower panel have 
identical start times and are temporally separated only by their respective exposure times of 2.9 and 
4.1 s. Given that the motion along the dotted line in Fig~\ref{fig:overlay} is about 0.2$\arcsec$ s\tsp{-1},
the longer XRT exposure cannot explain the offset. We suggest that XRT is seeing material 
that has flowed down the field lines and further equilibrated with coronal plasma, reaching an ionization 
state not visible in AIA. 

\begin{figure}[t]
 \epsscale{1}
 \plotone{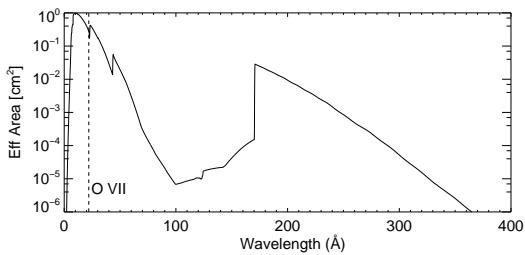}
 \caption{XRT/Al-mesh wavelength response showing the position of O VII lines (dashed line) that 
 	       likely dominate the cometary emission in this filter.}
 \label{fig:xrtres}
 \end{figure}

The likeliest candidate for this emission is O VII (22.1, 21.6 \AA{}), which lies near the peak sensitivity 
of the Al-mesh filter, the wavelength response of which is given in Figure~\ref{fig:xrtres}. 
Unlike AIA, XRT is a broadband instrument, and Al-mesh has some sensitivity to each of the lines 
contributing to AIA's EUV channels. 
Given the considerable signal exhibited by 
the 171 \AA{} band in particular, it was initially suspected that the faint XRT signal was also due to EUV 
emission. But the effective area at these lines is down by a factor of $\gtrsim$100 compared to O VII, and 
the lower ionization states cannot explain the observed offset. Such an offset 
would be expected from O VII because of the much longer ionization time compared to O VI
(see \S\ref{positions} for additional discussion of positional offsets between the emission in each 
passband). 

O VII is a line typical of a moderately active corona. Its temperature distribution function G(T) is peaked 
around 10\tsp{6.3} K with a FWHM of about 1.9 MK.
As will be described in \S\ref{dem}, 
we find a coronal temperature profile surrounding the comet that is peaked at 10\tsp{6.12} K, which 
would be high enough to generate O VII. It should be noted, however, that G(T) is calculated assuming 
ionization equilibrium. This assumption cannot be made for the cometary emission, but overlap 
between G(T) and the coronal temperature profile nonetheless provides some expectation of O VII formation. 


\section{Analysis}
\label{analysis}

\subsection{Coronal Differential Emission Measure}
\label{dem}

 
 \begin{deluxetable*}{l c c c c c c c c}
\tablecolumns{9} 
\tablewidth{0pc} 
\tablecaption{Measurement Results \& Outgassing Rates} 
\tablehead{ 
\colhead{Channel} & 
\colhead{Photons per Atom} & 
\colhead{Atoms per Photon} & 
\multicolumn{2}{c}{Cometary Flux\tablenotemark{a}} &
\multicolumn{2}{c}{Cometary Portion} &
\multicolumn{2}{c}{Outgassing Rate\tablenotemark{b}}
\\ 
\colhead{(\AA{})} & 
\colhead{at Earth} & 
\colhead{at Comet} & 
\multicolumn{2}{c}{log(DN s\tsp{-1})} &
\multicolumn{2}{c}{of Total Flux} &
\multicolumn{2}{c}{log($\mathsf{\dot{N}}$) [$\mathsf{N}$ s\tsp{-1}]}
\\
\colhead{} & 
\colhead{($\gamma$N\tsp{-1} )} & 
\colhead{log($4\pi{}\cdot$1AU\tsp{2}$\cdot{}\mathsf{N}\gamma\tsp{-1}$)} & 
\colhead{Global\tablenotemark{c}}  &
\colhead{Individual\tablenotemark{c}} &
\colhead{Global\tablenotemark{c}} &
\colhead{Individual\tablenotemark{c}} &
\colhead{Global\tablenotemark{c}} &
\colhead{Individual\tablenotemark{c}}
}
\startdata		
 1600 & 1.46   &   27.28     &  2.149 $\pm$ 35\% & 2.154 $\pm$ 18\%& 6.4\%& 23\% &   30.35 &   30.35  \\
 335\tablenotemark{d} & 0.0032   &   29.94  &  2.002 $\pm$ 47\% & 1.846 $\pm$ 50\%& 5.5\%& 7.5\% &  31.78 &   31.63  \\
 304 & 0.0021   &   30.13  &  2.399 $\pm$ 22\% & 2.302 $\pm$ 20\%& 9.9\%& 16\% &  32.74   &   32.64 \\
211  & 0.0180   &   29.19  &  3.364 $\pm$ 12\% & 3.290 $\pm$ 11\%& 3.2\%& 5.1\% &  32.60   &   32.52 \\
193 & 0.0920  &    28.49  & 4.077 $\pm$ 4.9\% & 4.037 $\pm$ 4.4\%& 4.3\%& 6.0\% &   32.57  &    32.53 \\
171 & 0.305    &  27.96    &  4.535 $\pm$ 1.4\% & 4.533 $\pm$ 1.4\%& 21\%& 21\% &  32.46  &    32.46\\
131\tablenotemark{d} & 0.0306   &   28.96  &  3.190 $\pm$ 4.4\% & 3.189 $\pm$ 4.4\%& 43\%& 43\% &  31.99    &  31.99 \\
\hline
Al Mesh\tablenotemark{e} &  \nodata &  \nodata & \multicolumn{2}{c}{2.576 $\pm$ 59\%}&  \multicolumn{2}{c}{3.1\%}  &   \multicolumn{2}{c}{\nodata}
\enddata
\tablenotetext{a}{Fluxes given in measured units. Photons =  
DN $\times$ Gain (e\tsp{-}DN\tsp{-1}) $\times$ 3.65 eV(e\tsp{-})\tsp{-1} $\times$ $\lambda$(hc)\tsp{-1}, where the CCD gain is 
$\sim$18 for the AIA telescopes and 59 for XRT. Taking $\lambda$ from the dominant lines in Table~\ref{tab:ions}, column 
4 converted to log($\gamma$ s\tsp{-1}) is: 3.064, 1.840, 
2.609, 3.404, 4.086, 4.495, 3.028, and 2.165. Values are approximate
because all contributing $\lambda$s are not considered. For this reason, we chose to list DN.}
\tablenotetext{b}{\ndot{O} for all channels except for 1600 \AA{}, which is \ndot{C}.}
\tablenotetext{c}{``Global" results taken from an image mask that encompasses all emission in all channels. ``Individual" results taken from 
			masks individually contoured to the emission in each channel.}
\tablenotetext{d}{The 131 and 335 \AA{} results may be underestimates. See \S\ref{131335} for details.}
\tablenotetext{e}{\ndot{O} cannot be calculated for O VII because the plasma does not \textit{quickly} ionize through this stage.}
\label{tab:flux}
\end{deluxetable*}


To analyze the plasma through which the comet is moving, we have computed a Differential 
Emission Measure (DEM) for a region surrounding the cometary emission. 
DEMs summarize coronal temperature and density structure 
by combining observations of several ions with various characteristic temperatures. This calculation 
assumes ionization equilibrium and so is not appropriate for the cometary emission itself. Rather, 
it provides 
a useful constraint on the conditions that Lovejoy experienced, which is important to our analysis 
for two reasons: 1) as an input to the outgassing calculations in \S\ref{outgassing}, and 2) for 
discussing the possibility of high-temperature lines like O VII, as in \S\ref{xrt}.

The SolarSoft\footnote{http://www.lmsal.com/solarsoft/}
routine $<$xrt\_dem\_iterative2$>$ was used to calculate the DEM using fluxes from 
the following bands: XRT/Al-mesh and AIA/94, 131, 171, 193, \& 211 \AA{}. This software was originally 
developed for XRT \citep{We04} and has since be adapted to incorporate AIA data \citep{Ch12}.

The 
approach uses forward-fitting to estimate a DEM that is then fed into the XRT and AIA filter response 
functions to predict fluxes. The final DEM is found by minimizing the $\chi{}$\tsp{2} between the 
predicted and observed emission. Monte Carlo simulations are also computed by randomly varying 
the observed intensities within their RMS noise; the spread of these simulations is then a gauge of the 
uncertainty (and its distribution) in each log(T) bin. 
Note that the inclusion of XRT data is particularly important for constraining 
the high-temperature component, which may be overestimated by DEM solutions that 
rely solely on AIA.

\begin{figure}[t]
 \epsscale{1}
 \plotone{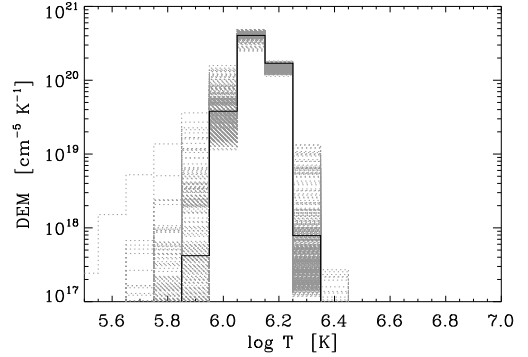}
 \caption{DEM distribution for the corona surrounding the comet at 00:46:12 UT. Solid represents the 
 best fit distribution, while the dotted gray segments indicate Monte Carlo simulations estimating 
 the uncertainty in each log(T) bin.}
 \label{fig:dem}
 \end{figure}

Input emission counts were taken from 
a ring immediately surrounding the comet region described in \S\ref{observations}. This ring covered 
an area of 2030 AIA pixels, and the resultant DEM distribution is displayed in Figure~\ref{fig:dem}.
We find the coronal environment surrounding 
the comet at 00:46:12 UT to be characterized by an average temperature of log(T) = 6.15 K, with 
contributions from plasma between log(T) $\approx$ 5.9 - 6.3 K. The total emission measure 
covering the full range of temperatures is 1.90\e{26} cm\tsp{-5}. 


\subsection{Outgassing Rates}
\label{outgassing}

As water sublimates from the nucleus, photodissociation of the H\tsb{2}O molecules produces 
oxygen atoms that are quickly ionized to charge states typical of the corona. Each O atom lingers 
in a given charge state for time (n\tsb{e}q\tsb{i})\tsp{-1}, where n\tsb{e} is the electron density and 
q\tsb{i} is the ionization rate for that ion. See Table~\ref{tab:ions} for q\tsb{i} and 
note that a typical coronal n\tsb{e} of 10\tsp{8} cm\tsp{-3} was assumed for the proceeding calculations. 
Given these values, a neutral O atom is excited to O I in $\sim$0.12 s.\footnote{This time 
is actually a bit shorter because charge exchange, which we have not accounted for, is a significant ionization mechanism for the 
lowest stage. Ionization of subsequent states is dominated by 
electron collisions. See \citet{BP12} for a detailed discussion of this.} The ionization rate for each subsequent 
state is 3--5x smaller, and so an ion persists in those states for correspondingly longer durations.

During its time in a given charge state, an atom is excited at rate n\tsb{e}q\tsb{ex} 
and produces q\tsb{ex}q\tsb{i}\tsp{-1} photons, where q\tsb{ex} is the excitation rate. From this 
and the instrument response to each ion, a total O outgassing rate in atoms per second (\ndot{O}) can be calculated. 
The calculation is the sum over spectral lines, $l$, of ions, $Z$: 

\begin{equation}
\textrm{count rate} = \sum_Z\sum_l  A_l \frac{q_{ex,l}}{q_{i,Z}} \ \textrm{\ndot{O}}/4\pi \textrm{D\tsp{2}} 
\end{equation} 

\noindent Where $A_l$ is the effective area of the AIA band for each line and the
distance D is 1 AU. 

Outgassing rates are listed alongside their corresponding fluxes in columns 8 and 9 of 
Table~\ref{tab:flux}. The former shows rates determined via fluxes extracted from a single (global) region that encompasses
all emission across all bands, and the latter shows rates determined via fluxes extracted from regions contoured 
to the shape of the emission in each channel. 
The second and third columns include important values calculated en route to the 
final outgassing rates. Column 2 lists the number of photons to be expected at Earth from a given 
ion before it reaches the next charge state, while 3 lists the number of atoms to be expected at 
the comet for each photon detected at Earth.
 For bands containing 
two ions (see Table~\ref{tab:ions}), the values quoted reflect the sum of both. A pivotal assumption 
behind these calculations is that the plasma is quickly ionized through each state. Since this 
is not true of O VII, an outgassing rate cannot be derived from the XRT observations. 

Across the EUV channels, we find an average \ndot{O} of 10\tsp{32.47} at 00:46:12 UT (column 9).
For comparison, SOHO/UVCS spectra taken at 6.8 \rsolar{} ($\sim$4.5 hrs before perihelion) suggest \ndot{H} = 10\tsp{32.48} 
\citep{Ra13}.
If all O and H atoms are assumed to come from H\tsb{2}O, then the UVCS result is down from AIA by about half in 
\ndot{H\tsb{2}O}. See 
\S\ref{discussion} for additional commentary on these results. 
A carbon outgassing rate can be similarly estimated from the 1600 \AA{} emission, and for this we 
find \ndot{C} = 10\tsp{30.35}. This is consistent with expectations based on the Si:O ratio 
of $\sim$0.05 observed by UVCS and the C:Si ratio of $\sim$0.1 reported for fellow Kreutz sungrazer C/2001 C2 \citep{Ci10}. Note 
that the rates we derive are somewhat dependent on temperature. For instance, if we were to have assumed 
log(T) = 6.25 K instead of 6.15 K, the resultant outgassing rates would be about a factor of 1.4 larger. 

The \ndot{O} values derived from the 304, 211, 193, and 171 \AA{} channels agree well, with a 
standard deviation that is about 18\% of their mean for the numbers listed in column 9. However, 
the rates found from the 335 and 131 \AA{} bands are not so consistent; about a factor of 5 separates 
the average of these results compared to that of the other EUV channels. 
Setting this discrepancy, which we will consider in the next section, aside for a moment, 
the variance in our results could be attributed to a number of factors:

For this analysis, we have assumed that all of the emission arises from the ions listed in 
Table~\ref{tab:ions}. As noted in \S\ref{euv}, \citet{BP12} found an appreciable contribution from Fe
to the 171 and 131 \AA{} bands that we have not accounted for ($\sim$40 \& 25\%, respectively). 
A modest Fe contribution to the 171 \AA{} band would indeed reduce the overall discrepancy somewhat. 
Likewise, an iron contribution to the 131 \AA{} channel would also bring that result closer the 335 \AA{} value, but 
whether or not this helps the overall variance depends on the issues discussed in \S\ref{131335}. It is also possible 
that additional lines unaccounted for in the 
CHIANTI database are contributing to the observed fluxes, and uncertainties in the calibration of the AIA response to the 
ions in Table~\ref{tab:ions} could be important. 
However, one of the most significant contributors to the variance in our results is likely atomic rate 
uncertainties, which are particularly high for EUV lines of the low ionization states being discussed 
here since it is not often that they must be considered. 

\subsubsection{131 and 335 \AA{} Discrepancy}
\label{131335}

The oxygen outgassing rates derived from the 131 and 335 \AA{} channels are down by a 
factor of $\sim$5 from those found for the other bands. O VI lines at 129--130 \AA{} 
dominate the predicted count rates for both channels, which are housed together 
in one of the four AIA telescopes. Each half of the telescope's aperture is coated for a particular 
channel and is illuminated at all times. Focal-plane filters are used to select the 
active channel, but the 335 \AA{} filter does not fully reject 131 \AA{} light \citep{Bo12}. This 
is generally a minor effect but is important here because the O III lines near 
335 \AA{} are much weaker than their O VI counterparts near 131 \AA{}. 

Interestingly, if the 131 \AA{} contribution to the 335 \AA{} response is \textit{not} included, 
the resultant log(\ndot{O}) is found to be 32.49. This is in good agreement with 
the other bands. If the 131 \AA{} contribution \textit{is} included, then the 335 and 131 \AA{} 
results agree fairly well with each other, particularly given the 
131 \AA{} Fe contribution that has not been included and the low 335 \AA{} signal-to-noise. 
Regardless of the 335 \AA{} response, 
the low 131 \AA{} outgassing result must be reconciled with those of the other bands 
since the atomic rates for O VI should be reliable. 

One way to account for this is to assume that, while most of the oxygen atoms progress all the way 
from O III through O VI, only about 1/3 of them are ionized from 
O VI to O VII, so that the intensity of the O VI line is only about 1/3 of the value assumed 
in column 2 of Table~\ref{tab:flux}. This is plausible in that we are looking at a set of 
images dominated by a short-lived outburst (see \S\ref{outbursts}), and the ionization time of O VI is 3$\times$ longer 
than that of O V, which is in turn 4$\times$ longer than that of O IV. 
O VI lines also contribute to the 171 \AA{} band \citep{BP12}, so \ndot{O} 
from 171 would have to be increased by perhaps 20\%, but this is offset by the Fe IX 
contribution that we have not included. 

If that interpretation is right, \ndot{O} around 10\tsp{32.5} matches all the bands to 
within a factor of $\sim$1.5. It implies that less of the oxygen has reached O VII than we 
had assumed, so a somewhat higher density would be required to account for the XRT count 
rate. A more complete model of the outgassing rate during the outburst and the time-dependent 
ionization would improve the estimates, but atomic rate uncertainties of $\sim$30\% and 
AIA calibration uncertainties limit the ultimate attainable accuracy. 

It should also be noted that a higher temperature due to thermalization of kinetic energy from cometary ions would 
decrease the O VI emissivity relative to that of the other ions somewhat, but not very strongly. 
However, this would increase the O VII emissivity and the XRT count rate dramatically.


\newpage

   \begin{figure}[h]
 \epsscale{1}
 \plotone{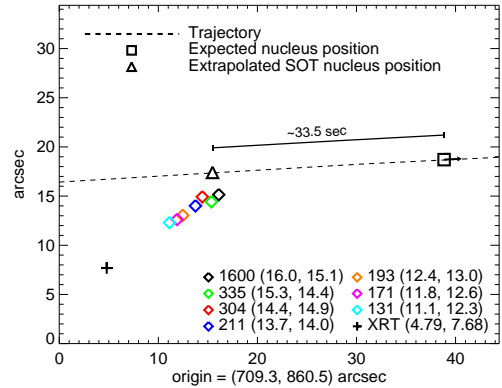}
 \caption{Westernmost, leading edge of the tail in each band surrounding the 171 \AA{} observation 
 at 00:46:12 UT. Trajectory (dotted) and expected nucleus position (square) are taken from JPL 
 orbit solution \#58.
 Extrapolated SOT position (triangle) corresponds to the delay observed by that 
 instrument during the ingress. Parenthetical values in the lower-right indicate positions  
in arcsec offset from the origin that in turn indicates distance from Sun-center. 
Note that these positions are shifted based on observation times and the comet's trajectory to estimate 
locations at 00:46:12 UT.}
 \label{fig:tips}
 \end{figure}

\subsection{Positional Comparison}
\label{positions}

The westernmost, leading tip of the emission in each channel was determined manually, and the results 
are displayed in Figure~\ref{fig:tips}. Trajectory information is derived from the JPL orbit 
solution \#58 and extracted using the HORIZONS ephemeris 
generator\footnote{http://ssd.jpl.nasa.gov/?horizons}. 
Since minor differences in the tip 
positions can be informative, a few considerations are particularly important. Translational, rotational, 
and scaling offsets between the four AIA telescopes are accounted for using the $<$aia\_prep$>$ software 
in SolarSoft. We have used the latest version of this code, which includes refinements to the 
the telescope offsets resulting from the 2012 Transit of Venus observations. 
The XRT observations were scaled to AIA's pixel size, rotated to match, and carefully 
aligned to the limb and on-disk features.  

Because the comet is traveling at $\sim$0.7\arcsec s\tsp{-1} and the observations 
are spread over 12 s, it is necessary to shift the tip positions 
found after co-alignment to account for the different observation and exposure times. Shifts were 
determined using the temporal offsets between the images and the expected motion of the comet 
according to the JPL ephemeris, which is given as a dotted line in Figure~\ref{fig:tips}. Note that 
the resolution of AIA is 0.6\arcsec px\tsp{-1} and the alignment between the telescopes is likely 
good to within 1-2 pixels. Thus, the $\sim$0.8\arcsec offset between the 131 (light blue) and 171 \AA{} (pink) positions, for instance, 
is close to the uncertainty from resolution and spatial alignment. 

The resultant 
positions arrange themselves almost exactly as would be expected from emission dominated by 
the ions listed in Table~\ref{tab:ions}. An exception to this is the 335 \AA{} position. This appears 
consistent with emission from O III, but as we describe in \S\ref{131335}, we expect this channel 
to be dominated by O VI. This is a curious but not overly concerning result because, as is 
evident in Fig~\ref{fig:images}, the 335 \AA{} band has very low signal-to-noise, making a precise 
measurement of the tip position difficult. 

C IV (1600 \AA{}) is the first to appear, taking $\sim$1.7 s to reach the fourth ionization stage 
after being sublimated, given a coronal electron density of 10\tsp{8} cm\tsp{-3} and log(T) = 6.15 K. The separation 
from the nucleus should be $\sim$0.25$\times$ the length of the 1600 \AA{} tail, or about 3.25\arcsec. 
Along a linear fit through tip positions from each passband, the distance between the 1600 \AA{} 
tip and the orbital path is $\sim$3.5\arcsec. We therefore see about a 28 s delay between the 
expected nucleus position and that suggested by the AIA observations. 

The Solar Optical Telescope 
(SOT) aboard Hinode also observed the comet during its ingress on the opposite side of the solar disk. 
The observed position was precisely along the expected orbital path but $\sim$33.5 s behind the 
predicted position. The extrapolated SOT position based on this delay is denoted by a triangle 
in Figure~\ref{fig:tips}, and one can see much better agreement between this position and 
the expectation from AIA. Ground-based observations well outside perihelion were used to construct 
the ephemeris, so the solution very near the Sun may suffer a small timing error as a result. It should be also 
be noted that the orbit determination for Comet Lovejoy is still ongoing, so agreement between the ephemeris and 
the solar telescope observations will likely improve.

The distance (d) between the tips seen in O III and VI (304 and 131 \AA) can also be used to 
estimate the density of the emitting plasma:
\begin{equation}
\mathrm{d} \approx V\tsb{comet} * \mathrm{n}\tsb{e}\tsp{-1}(\mathrm{q}\tsb{III}\tsp{-1} + \mathrm{q}\tsb{IV}\tsp{-1} + \mathrm{q}\tsb{V}\tsp{-1})
\end{equation}
\noindent Where 
the ionization rate coefficients q\tsb{i} are those given in Table~\ref{tab:ions}. 
Following from this, we find an electron density of $\sim$1.4\e{8} \cm{}. 

 
 \subsection{Opening Angle}
 \label{angle}
 
The opening angle of the tail is informative of the interplay 
between outgassed material and the local magnetic field. 
It can be well-measured in
the 171 \AA{} images, which exhibit the highest signal-to-noise. 
To do this, we manually determine a vector that roughly 
bisects the emission and is perpendicular to the striations 
along which cometary material is collected by the magnetic 
field (further discussion of these striations can be found in 
\S\ref{morph}). The emission is then divided into slices 
perpendicular to this vector, a Gaussian is fit to the 
emission along each slice, and the width of the comet 
at each location is determined by the distance between the 
3-$\sigma$ levels on either side of the Gaussian distribution.
From this, we find the opening angle of the 171 \AA{} emission 
at 00:46:12 UT to be 12.8$^\circ$.

For Lyman-alpha images
of sungrazing comets at larger heights, the
interpretation of the opening angle is straightforward
\citep{Be05}:  Hydrogen
atoms from the dissociation of cometary water exchange
electrons with coronal protons, producing a cloud of
neutrals that have the temperature and velocity of
the local coronal environment.  The cloud expands at the thermal speed,
producing a cone with opening angle tan\tsp{-1}($V$\tsb{thermal}/$V$\tsb{comet}).

The case of Comet Lovejoy is more complex.  An oxygen
atom is moving at $V$\tsb{comet} upon release from the nucleus.
As it becomes ionized
and begins to interact with the coronal magnetic field,
the O atom behaves like a pickup ion in the solar wind \citep{GG01}.  The velocity
component perpendicular to the field becomes a gyration
speed, $V_\bot$, and the parallel component 
is conserved as $V_\|$.  Since all the ions have
nearly the same speed, they form a tight ring in velocity
space.  This ring distribution is very unstable and 
quickly scatters into a bispherical shell distribution
with components of two intersecting shells centered
at $V_\| \pm V_A$, where $V_A$ is the Alfv\'{e}n speed \citep{WZ94}.

Thus, after deprojection from plane-of-the-sky 
to true values, the angle between the comet's path and
the center of its tail gives $V_\|$ / $V$\tsb{comet}. And the
opening angle of the tail gives the width of the bispherical
distribution, which depends on $V_A$, $V$\tsb{comet}, and
the angle between $V$\tsb{comet} and \textbf{B}.  That
analysis is beyond the scope of this paper because it
requires knowledge of the \textbf{B} field direction and because
the bispherical shell may further scatter, exchanging
energy and momentum with the coronal plasma.


 \subsection{Time Evolution}
 \label{time}
 
 
   \begin{figure*}[t]
 \epsscale{1}
 \plotone{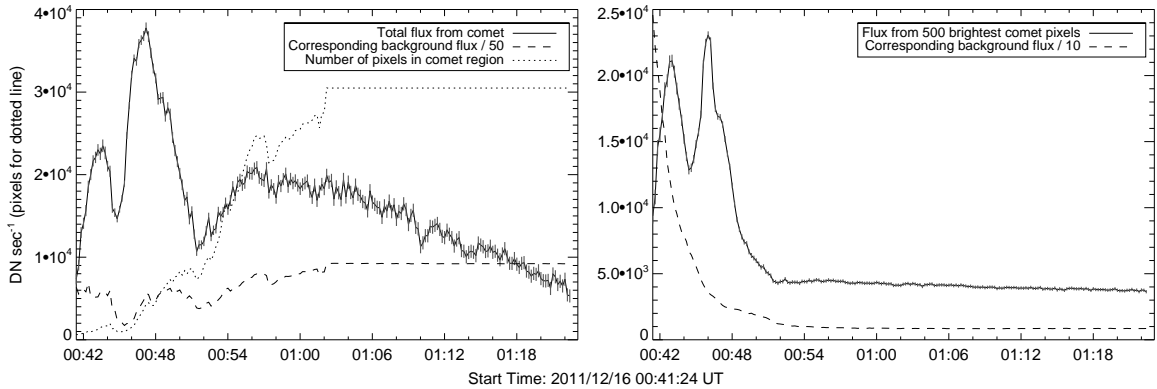}
 \caption{\textit{Left}: Total 171 \AA{} cometary flux (solid) alongside corresponding background 
 emission divided by 50 (dashed) and area of emission region in pixels (dotted). 
 \textit{Right}: Total 171 \AA{} flux from the 500 brightest comet 
 pixels (solid) alongside flux from corresponding background pixels divided by 10 (dashed). Perihelion 
 occurred $\sim$25 min prior to the plots' start time, and the leading edge of the tail 
 exits the FOV at 00:56:00 UT. 00:41:00 and 00:56:00 UT correspond to heights above the photosphere of 
 0.46 and 0.78 \rsolar{}, respectively.}
 \label{fig:fluxtime}
 \end{figure*}
 
This section is intended as a prelude to future work that will cover the time evolution of Comet Lovejoy in more detail. 
Here we make some qualitative remarks about morphological developments during the egress, followed by 
a quantitative analysis of the AIA/171 flux versus time. 

\subsubsection{Morphological Developments}
\label{morph}

A number of interesting features in this dataset bear mentioning. We do so here with references to 
the positions labeled in Figure~\ref{fig:rfilter}, but 
readers would be best served by \href{http://aia.cfa.harvard.edu/movies/comet_lovejoy_egress_171.mp4}{downloading the corresponding movie} available in our online material. 
Traveling roughly horizontally  
with respect to the FOV, the comet emerged from behind the limb $\sim$25 min after perihelion. 
Interaction between shed cometary material and the local 
magnetic field is immediately apparent, as a parcel of newly-formed plasma can be seen traveling away from the Sun 
along a radial field system near position 1. Ionized cometary material will 
tend to spiral along field lines with a speed that depends on $V$\tsb{comet} and the angle between $V$\tsb{comet} and \textbf{B}. 
For this initial interaction, the component of $V$\tsb{comet} along \textbf{B} appears to be greatest, so we observe the swiftest motion 
of cometary material along field lines here. Similar, but slower, motions can be observed near positions 
4 and 5 in a southwestern direction and in a northwestern direction near position 6. 

In position 2, we see the tail stretched out in a slight, inverted u-shape. This 
appears to be the comet skimming overtop of underlying loops, but modeling of the magnetic field will be 
needed to understand this. The same is true for the vertical striations that become most pronounced after position 4, which 
can be interpreted as newly formed plasma being arranged along field lines. As was pointed out by \citet{BP12}, such 
striations are also visible during the ingress. Other groups are working on magnetohydrodynamic (MHD) modeling of 
this system, and it will be interesting to see how closely correlated these striations are with the model field predictions.

Position 6 shows the last frame for which the leading edge of the EUV emission can be traced before it has left 
the FOV. This occurs at about 00:56:00 UT when the comet is $\sim$\rsolar{0.56} above the photosphere.
Emission visible in the radial filter movie lingers along the striations at this position for another $\sim$40 minutes. 
The motion of these striations during this period is very interesting, with the roughly vertical threads pulling apart and 
moving together in a manner reminiscent of aurorae on Earth.

\subsubsection{Flux vs. Time}
\label{fluxvtime}

Figure~\ref{fig:fluxtime} quantifies the 171 \AA{} brightness over a 40-minute period from 00:41:24 
through 01:22:24 UT. This begins about 25 minutes after perihelion, corresponding to a height of $\sim$\rsolar{0.46} 
above the photosphere, and ends $\sim$26 minutes after the leading edge of the tail
 has left the FOV.
The solid line in the 
left panel shows the total flux extracted after background subtraction from a region contoured to fit the cometary 
emission. The dashed line 
indicates the total background flux from the same region, and the dotted line shows the area of this region in pixels.

The background subtraction and region selection for this analysis was done in the same manner as described in 
\S\ref{observations} with some modifications. 
Since the tail's extent and position changes from frame-to-frame, a new ``comet region" (mask) 
must be selected for each image. To make the evolution of this region as continuous as possible, the mask for each 
frame was taken from the previous image and adjusted to fit the new extent. The dotted line in the left panel of 
Fig~\ref{fig:fluxtime} indicates the area of this mask,  showing the dramatic rise of the emission's extent as time progresses. 
Note that the background flux in the left panel (dashed line) is largely dependent on the area indicated by the dotted line. 
Though coronal emission falls off exponentially 
with radial distance, the size of the comet mask increases nearly as fast, yielding a relatively constant background flux 
when integrated over the entire region. 

To display these data in a manner independent of the mask size, the right panel of Fig~\ref{fig:fluxtime} shows the 
cometary flux integrated over only the 500 
brightest pixels in the the selected region (solid line). The dashed line again shows the background flux integrated 
over the same pixels used for the solid line. Since a constant area is being used, the background curve drops off as would be expected of the corona.  

As previously noted, the leading edge of the tail can be seen exiting the FOV at about 00:56:00 UT, 
which corresponds to a local maximum in the comet's flux as shown in the left panel of Fig~\ref{fig:fluxtime}. After this 
time, lingering emission drifts about for many minutes confined to a region somewhat larger than the extent indicated 
by position 6 in Fig~\ref{fig:rfilter}. The frame-by-frame evolution of the comet mask is discontinued after 01:02:00 UT in 
favor of a constant region; the flat portion of the dotted line in the left panel of Fig~\ref{fig:fluxtime} results from this. 

We find the AIA/171 brightness to be characterized by three distinct peaks and leave their interpretation to the proceeding 
section, where we discuss the profiles shown in Fig~\ref{fig:fluxtime} in the context of the outburst model proposed 
by \citet{SC12}. The total integrated emission under the flux vs. time plot in Fig~\ref{fig:fluxtime} is  
log(I) = 7.609 DN, which implies to a total outgassing of 10\tsp{35.53} oxygen atoms based on the 171 \AA{} 
production rates listed in Table~\ref{tab:flux}. Note that in applying the Table~\ref{tab:flux} rates to the integrated flux, we 
are in effect assuming that the corona is of uniform temperature and density throughout the AIA FOV. 
If entirely from water, our total outgassing result translates 
to a mass loss of $\sim$1.0\e{13} g. For comparison, \citet{SC12} used dust observations to estimate a residual mass of the nucleus 
of $\sim$1\e{12} g after its destruction 1.6 days post-perihelion at \rsolar{31}. We will discuss the implications of our 
mass loss estimate on the expected size of the nucleus in \S\ref{nucleus}. 

	
\section{Discussion}
\label{discussion}


\subsection{Outbursts}
\label{outbursts}

\citet{SC12} used preliminary results from this study, presented at the 2012 Summer AAS meeting, to test their model for the 
process by which Comet Lovejoy's 
nucleus may have been destroyed, which was touched on in \S\ref{intro}. In summary, they contend that the nuclei of sungrazers 
like this one are principally destroyed through successive, large fragmentation events, rather than by steady outgassing. As 
heat propagates into the interior of the nucleus, pockets of water ice explode, calving off large chunks of material in outburst events. 
This results from an exothermic reaction triggered by the crystallization of amorphous ice at a critical temperature of $\sim$130 K \citep{Sc89}.
Such an interpretation is significant to our general understanding of cometary compositions because it implies that the nucleus 
has a tensile strength far greater than what would be implied by the canonical ``rubble pile" model. 

In our AAS poster, we performed the outgassing calculations described in \S\ref{outgassing} at three times (2012/11/16 00:44:12, 00:46:12, 
and 00:48:12 UT), and these results began to trace out the second of the two sharp peaks in Fig~\ref{fig:fluxtime}. \citeauthor{SC12} note that 
these rates could not be sustained by a sub-kilometer nucleus for very long. Instead they suggest that without drastically increasing 
the expected size of the nucleus, our results could only be explained in the context of an outburst, perhaps one of the four post-perihelion 
events that they suspect ultimately led to the total destruction of the nucleus. 

With the fuller analysis presented in \S\ref{fluxvtime}, we see that the 171 \AA{} intensity is characterized by three distinct peaks, 
the first two of which are sharp spikes that occur 
while the emission is confined to a relatively compact region spanning less than $\sim$50$\arcsec$. These are centered near 
00:43:00 and 00:47:00 UT, respectively, and we find them to be consistent with the outburst model proposed by 
\citet{SC12}. It is apparent from the two-peaked profile that material was expelled from the nucleus in distinct phases, but whether 
or not this implies explosions in distinct pockets of water ice is unclear. 
Further evidence for the outburst scenario may be found in the relative contributions of O and Fe to the observed emission. 
This would be informative of the source of neutral oxygen, which may be either water ice (as we have assumed) or dust grains 
(as proposed by \citeauthor{SC12}). 
We discuss these possibilities in the next section. 

Though the overall intensity of the third peak at 00:56:00 UT is
comparable to the first two, the emission is far more diffuse and the rise more gradual than the earlier peaks, 
which are dominated by dense collections of very bright pixels. If the first two spikes are the result of outbursts that 
ejected large volumes of material from the 
nucleus, then we speculate that third peak may arise from more gradual outgassing through newly opened vents. 
Since the front of the tail leaves the FOV at the time of the third peak, we may also be seeing the beginning of a 
much longer rise phase. Even if this were true, the nature of this event does not appear to be quite the same as the first two because 
dense concentrations of material are not observed. 

There is also a question of why the emission lingers 
for so long near the FOV edge when this behavior is not found to the same extent earlier. We suspect this to 
be primarily a result of density contrast. The coronal density is likely much less at this location
than closer in, so 
the time it takes for the cometary material to diffuse into the background is much longer. This is coupled with the  
fact that a lower density and likely temperature will increase ionization times, so we see material progressing into 
and leaving the charge states observable at 171 \AA{} more slowly.  


\subsection{Oxygen Source}
\label{oxygen}

\citet{SC12} have suggested that much of the oxygen seen in the AIA images originates from the 
vaporization of cometary dust grains rather than sublimation of water ice. From an observational 
standpoint, the two possibilities can be distinguished if the very high abundances of Si, Mg, and Fe 
relative to O in dust imply significant emission in some of the AIA channels. We take the Mg:Si and 
Fe:Si abundance ratios to be 0.69, as did \citet{BP12} based on the abundances in Comet Halley \citep{De88}. 
Assuming that the Si:O ratio is 0.5, we used Version 7 of CHIANTI \citep{chianti,chianti7} to compute contributions of 
the refractory elements. This ratio is what we would expect if the dominant oxygen source is dust 
grains; note that it is an order of magnitude higher than what we reference in \S\ref{outgassing}.

Si would produce only 10\% increases in the AIA bands except for the 94 \AA{} band, where it would 
provide a large fractional increase but still leave the emission at an undetectable level. Mg would increase 
the brightness in the 304 and 335 \AA{} bands by about a factor of 3, which would aggravate the disagreement 
among the channels. It would also make the 304 \AA{} image appear longer and more displaced from 
the comet nucleus than observed. 

If we assume that n\tsb{e}t is large enough to ionize O through O VI, it will also ionize Fe through Fe IX. In that 
case, Fe VIII and Fe IX make substantial contributions to a few bands, particularly at 171 \AA{}. A modest 
contribution reduces the discrepancy among the channels, but if the oxygen comes from dust grains, the expected iron
levels would increase the 171 \AA{} brightness by a factor of six, pushing it out of line with the other channels. 
An Fe:O ratio of about 0.05 seems to give the best agreement, suggesting a grain-to-water ratio of about 1/6. 
This is in line with the Si:H ratio obtained from the UVCS spectrum at \rsolar{2} during the ingress \citep{Ra13}.

It is important to note that this discussion pertains to the frames we have analyzed in detail. These were 
centered around 00:46:12 UT, which lies near the peak intensity indicated by Fig~\ref{fig:fluxtime}. 
We have suggested that this peak is consistent with the outburst scenario proposed by \citet{SC12}.
If true, then the intensity spike results from the explosion of an interior pocket of water ice that blew off a 
chunk of the nucleus. \citeauthor{SC12} point to dust grains as the primary source of neutral oxygen because 
they suspect most surface ice to have already been sublimated away by the time of these observations, with 
the only significant source of H\tsb{2}O left being these interior reservoirs. 

If this type of outburst is manifested in the AIA data, then presumably the ejecta contains high 
concentrations of water, which would be consistent with our results. It may also be true that when not in an 
outburst phase, the primary source of neutral oxygen is indeed dust grains. As we have just described, this 
could be determined by the relative contribution of Fe implied by outgassing rates derived near a minimum 
in the overall intensity versus that found for the maximum period this paper has focused on. If dust was found 
to be the dominant source of neutral O outside of the peak phase, this would also strengthen the notion 
that the intensity spike results from an explosion of interior water ice. These questions will be addressed in a followup study that 
will cover time evolution in more detail. 


\subsection{Comparison with C/2011 N3}
\label{n3}

As has been previously mentioned, the first sungrazing comet observed by AIA came a 
few months prior to Lovejoy when fellow Kreutz member C/2011 N3 burned up midway through its 
transit of the solar disk. These results were published by \citet{Sc12}, which was discussed briefly in \S\ref{intro}.
As with Lovejoy, emission from C/2011 N3 was detected in all of AIA's EUV channels except at 94 \AA{}. 

\citeauthor{Sc12} also estimated the mass loss of C/2011 N3 but did so in a very different manner than through 
the outgassing calculations presented here. They measured the deceleration of material lost from the nucleus 
and coupled this with the comet's orbital trajectory and approximate knowledge of the coronal plasma density to 
estimate a mass loss rate of (0.01-1)\e{8} g s\tsp{-1} over the duration of the comet's visibility to AIA. This amounted to a 
total mass loss of $\sim$6\e{8} to 6\e{10} g.

In 
\S\ref{outgassing}, we estimate an average \ndot{O} of 10\tsp{32.5} across the AIA channels for the C/2011 W3 
at 00:46:12 UT. Assuming all this comes from water, 
we have a mass loss rate of $\sim$9.5\e{9} g s\tsp{-1}
(see \S\ref{oxygen} for discussion on the source of neutral O). 
Note that this time lies near the peak intensity  
indicated by Fig~\ref{fig:fluxtime}. Aside from the first few frames, for which some of the emission may be 
hidden behind the solar disk, the minimum intensity observed before the leading edge of the 171 \AA{} emission 
exited the FOV occurred at 00:52:24 UT and was down by a factor of $\sim$3.2x from 46:12. For this time, we 
estimate \ndot{O} to be 10\tsp{31.95}, which translates to a mass lost rate of 2.7\e{9} g s\tsp{-1}. As noted in 
\S\ref{fluxvtime}, our total mass loss 
is estimated at $\sim$10\tsp{13} g. 

Our results are obviously considerably higher than what was reported for C/2011 N3, which was to be expected 
given that Lovejoy was a considerably larger comet. \citeauthor{Sc12} estimated that the nucleus of C/2011 N3 was 
between 10 and 50 meters during the AIA visibility, while Lovejoy was estimated at somewhat less than 1 km 
upon its solar approach \citep{Gu12}. Initial estimates from SOHO/UVCS spectra point to a diameter of 400 m by the time 
the comet reached \rsolar{6.8} pre-perihelion \citep{Ra13}. So at the time of our observations, Lovejoy's nucleus was likely at least 
an order of magnitude larger in diameter than C/2011 N3. Given this, a peak mass loss rate that is a bit 
more than an order of magnitude larger than what was derived for C/2011 N3 from a very different method is an encouraging result. 
We discuss the size of the nucleus implied by our own results in the next section. 

\subsection{Size of the Nucleus}
\label{nucleus}

The size of the nucleus at the start of our dataset can be inferred from the total mass loss of $\sim$10\tsp{13} g that we 
estimate in \S\ref{fluxvtime}. To do this, we must first make an assumption about bulk density. 
Like \citet{SC12}, we take the bulk density to be 0.4 g cm\tsp{-3} based on the Deep Impact results from 
Comet 9P/Tempel 1 \citep{Ri07}. This result ranged from 0.2--1 g cm\tsp{-3} with a preferred value of 0.4, and how well 
it applies to Comet Lovejoy is uncertain. See the review by \citet{WL08} for a conglomeration of various cometary density 
estimates, which suggested a ``best" average value of 0.6 g cm\tsp{-3}.
At 0.4 g cm\tsp{-3}, our mass loss translates to a sphere of diameter $\sim$363 m. 

\citet{SC12} estimate that the nucleus was between 150 and 200 m in diameter at the time of its destruction 1.6 days after perihelion. 
Allowing for some outgassing after the comet left the AIA field-of-view, we find that the nucleus was at least 400 m in diameter 
when our observations begin, 25 minutes post-perihelion. 
Our mass loss estimate then corresponds to the erosion of a $\sim$73 m layer from the surface, 
leaving the nucleus with a diameter of $\sim$254 m when it exits the AIA FOV. At this size, another 
$\sim$(1.7--2.7)\e{12} g of material could be lost before reaching the 150--200 m diameter found by 
\citeauthor{SC12}. 

Considering the unknown amount of mass 
lost behind the Sun and on ingress, the nucleus may then have been around 600 m 
in diameter before the encounter. This would allow for just over 3 times the total 
outgassing derived here (3\e{13} g) to be sustained before the start of our 
observations. Note that this estimate is somewhat larger than the 400 m 
preliminary estimate from SOHO/UVCS at \rsolar{6.8} pre-perihelion \citep{Ra13}.


\section{Conclusion}
\label{conclusion}

We have explored the EUV and X-ray emission observed toward Comet Lovejoy using 
post-perihelion observations from AIA and XRT. Our results are summarized by the following bullet points:

\begin{itemize}

	\item{We find the emission to be dominated by oxygen ions
		with some contribution from iron, as proposed by \citet{BP12}. 
		These ions 
		form as neutrals from the comet sublimate, dissociate, and ionize toward equilibrium with 
		coronal plasma. We also note that, unlike the other channels, carbon dominates the 1600 \AA{} band. 
		[\S\ref{observations}]}
	
	\item{O III through VI are observed in the AIA EUV bands, along with C IV in the 1600 \AA{} UV channel. 
		These findings are based on cometary abundances, the lines listed in CHIANTI, and the effective areas of 
		the AIA filters. [\S\ref{euv}]}
		
	\item{O VII is observed by XRT. This conclusion is based on the spatial offset between 
		the tail observed in XRT and AIA, the wavelength 
		response of the Al-mesh filter, and expectations based on the coronal environment. [\S\ref{xrt}]}
		
	\item{A DEM analysis of the background corona finds Lovejoy's environment at 00:46:12 UT to be 
		characterized by an average log(T) = 6.15 K, with a total emission measure of 1.90\e{26} cm\tsp{-5}. [\S\ref{dem}]}
		
	\item{We find an average outgassing rate (\ndot{O}) of 10\tsp{32.47} across the EUV channels at 00:46:12 UT. 
		This calculation is a function of the coronal temperature and electron density, the ionization and excitation 
		rates for each species, and the instrument response to each ion. [\S\ref{outgassing}]}
		
	\item{From the 1600 \AA{} channel, we find \ndot{C} = 10\tsp{30.35}. This is consistent with expectations 
		based on the Si:O and C:Si ratios found by UVCS for another Kreutz sungrazer. [\S\ref{outgassing}]}
		
	\item{A positional comparison of the leading edge of the tail in each channel further suggests the emission 
		is dominated by the species we identify. It also suggests that the location of the nucleus is $\sim$28 s 
		delayed from expectations based on JPL orbit solution \#58. [\S\ref{positions}]}
		
	\item{n\tsb{e} $\approx $ 1.4\e{8} cm\tsp{-3} is found for the cometary plasma based on the distance between the 
		O III and VI (304 and 131 \AA{}) tips. [\S\ref{positions}]}
		
	\item{An opening angle of 12.8$^\circ$ is found for the 171 \AA{} tail at 00:46:12 UT. We note that 
		this angle differs from that of Ly-$\alpha$ tails 
		because O ions at low heights will behave like pickup ions in the solar wind, forming a complex 
		distribution that depends on $V$\tsb{comet}, $V_A$, and \textbf{B}. [\S\ref{angle}]}
		
	\item{We quantify the AIA/171 brightness and find it to be characterized by 3 distinct peaks in intensity, with 
		a factor of $\sim$3.5 separating the high and low values observed while the leading edge of the tail 
		remained in AIA's FOV. [\S\ref{time}]}
		
	\item{Integrating under the AIA/171 flux vs. time plot, we find log(I) = 7.609 DN. This 
		translates to 10\tsp{35.53} oxygen atoms and a mass loss of $\sim$10\tsp{13} g, assuming that the oxygen comes 
		from water and the background corona is of uniform temperature and density. [\S\ref{time}]}
	
	\item{We find the AIA/171 flux vs. time results to be consistent with the outburst interpretation of our 
		preliminary results by \citet{SC12}. [\S\ref{outbursts}]}
		
	\item{We dispute the notion presented by \citet{SC12} that the O ions observed around 00:46:12 UT arise 
		principally from vaporized dust grains rather than sublimated water. We base this on 
		the relative outgassing rates found, noting that the 171 \AA{} brightness 
		should be much greater if the O atoms come from dust because of the much higher Fe contribution 
		implied. [\S\ref{oxygen}]}
		
	\item{Comparing our results to those of \citet{Sc12} for C/2011 N3,
		 we find a peak mass loss rate 
		that is over an order of magnitude greater for Comet Lovejoy (9.5\e{9} vs. 1\e{8} g s\tsp{-1}), 
		which is consistent with expectations based on Lovejoy's much larger size. [\S\ref{n3}] }
		
	\item{Based on our total mass loss estimate of 10\tsp{13} g, we suggest that the nucleus 
		was at least 400 m in diameter at the start of our observations, 25 min after perihelion. At a bulk 
		density of 0.4 g cm\tsp{-3}, this is equivalent to the erosion of a 73 m surface layer, leaving 
		the nucleus at $\sim$254 m upon exiting the AIA FOV. [\S\ref{nucleus}]}

\end{itemize}

The AIA and XRT observations of Comet Lovejoy have provided an exciting new entry into the study of sungrazing comets. 
Several groups continue to work on this dataset, and a followup to this study will be prepared in the coming year. AIA and XRT are 
also poised to obtain similar images for the upcoming perihelion passage of Comet 2012 S1 (ISON), which
appears to be 5 to 10 times larger than Lovejoy and will 
reach its closet approach on November 28th, 2013. 

	
\acknowledgements
 
Support for this work was provided by the Smithsonian Astrophysical Observatory (SAO) via 
funding from Hinode/XRT through grant NNM07AB07C and SDO/AIA through grant SP02H1701R. 
Hinode is a Japanese mission 
developed and launched by ISAS/JAXA, with NAOJ as domestic partner and NASA and STFC 
(UK) as international partners. It is operated by these agencies in cooperation with ESA and 
the NSC (Norway). SDO is a NASA satellite, and the AIA instrument team is led by Lockheed 
Martin, with SAO as the major subcontractor.  
We gratefully acknowledge Ted Tarbell for his work in aligning the SOT and SDO data, which 
was important for Figure~\ref{fig:tips}. 
P.I.M. thanks the SSXG members at SAO for being characteristically helpful and welcoming. 
\linebreak \\
{\it Facilities:} \facility{Hinode (XRT)}; \facility{SDO (AIA)}


\end{document}